\documentclass[runningheads]{llncs}
\usepackage{graphicx}
\usepackage{makecell}
\usepackage{caption}
\usepackage{subcaption}

\usepackage{hyperref}

\makeatletter
\newcommand{\printfnsymbol}[1]{%
  \textsuperscript{\@fnsymbol{#1}}%
}
\makeatother
\begin{document}

\title{Weakly Unsupervised Domain Adaptation for Vestibular Schwannoma Segmentation}


\author{Shahad Hardan \thanks{These authors contributed equally} \and
Hussain Alasmawi \printfnsymbol{1} \and
Xiangjian Hou \and Mohammad Yaqub}
\authorrunning{Hardan et al.}

\institute{Mohamed bin Zayed University of Artificial Intelligence, Abu Dhabi, UAE \\
\email{\{shahad.hardan, hussain.alasmawi, xiangjian.hou, mohammad.yaqub\}@mbzuai.ac.ae}\\
\url{https://mbzuai-biomedia.com/biomedia/}}
\maketitle              
\begin{abstract}
Vestibular schwannoma (VS) is a non-cancerous tumor located next to the ear that can cause hearing loss. Most brain MRI images acquired from patients are contrast-enhanced T1 (ceT1), with a growing interest in high-resolution T2 images (hrT2) to replace ceT1, which involves the use of a contrast agent. As hrT2 images are currently scarce, it is less likely to train robust machine learning models to segment VS or other brain structures. In this work, we propose a weakly supervised machine learning approach that learns from only ceT1 scans and adapts to segment two structures from hrT2 scans: the VS and the cochlea from the crossMoDA dataset. Our model 1) generates fake hrT2 scans from ceT1 images and segmentation masks, 2) is trained using the fake hrT2 scans, 3) predicts the augmented real hrT2 scans, and 4) is retrained again using both the fake and real hrT2. The final result of this model has been computed on an unseen testing dataset provided by the 2022 crossMoDA challenge organizers. The mean dice score and average symmetric surface distance (ASSD) are 0.78 and 0.46, respectively. The predicted segmentation masks achieved a dice score of 0.83 and an ASSD of 0.56 on the VS, and a dice score of 0.74 and an ASSD of 0.35 on the cochleas. Among the 2022 crossMoDA challenge participants, our method was ranked the $8^{th}$. 

\end{abstract}

\section{Introduction}
Deep learning (DL) is becoming more popular due to its practicality and its ability to outperform experts in various applications, such as in the medical field. However, unlike humans who can adapt and learn new experiences from different existing ones, DL models are sensitive to the settings they are trained in and do not implicitly adapt to these unseen settings. For instance, changes due to different scanners, image acquisition protocols, and medical centers are among forms of variability \cite{Van_Transfer}. Domain Adaptation (DA) is a field in machine learning (ML) that deals with the distribution changes between different data. The cross-Modality Domain Adaptation (crossMoDA) challenge \cite{crossMoDA} introduced the first multi-class benchmark for unsupervised cross-modality DA. The challenge consists of two tasks: the segmentation of two structures in the hrT2 scans, and the classification of hrT2 images with vestibular schwannoma (VS) according to the Koos grade. Our work is focused on the first, which involves using contrast-enhanced T1 (ceT1) MRI as a source domain and high-resolution T2 (hrT2) MRI as a target domain to segment two objects: the VS and the cochleas. Vestibular schwannoma is a benign tumor in the brain that, in case of growth, affects the hearing nerves. As an intervention, open surgery or radiosurgery is performed to cure it. These operations require information about the volume and the exact location of the tumor \cite{crossMoDA}. Therefore, accurate segmentation of the relevant anatomy helps plan the operation properly and consequently increases the chance of patients' recovery. 
\section{Related Work}
Several studies tackled domain adaptation in the medical imaging field. However, they are mostly private, small, and aim for binary image segmentation,
unlike the crossMoDA dataset \cite{crossMoDA}. The challenge started in 2021, providing a total of 242 training images from the two domains (ceT1 and hrT2). In 2021, the winning model achieved a dice score of 0.857 for VS and 0.844 for the cochleas \cite{first-winner}. They applied CycleGAN for domain translation, then used the fakeT2 images to train the nnUNet. After that, they inferred pseudo-labels of the real hrT2, which are used to retrain the model. The second 2021 ranking model used nnUNet while applying the approaches of pixel alignment and self-training \cite{second-winner}. They generated fake hrT2 images using NiceGAN and achieved a mean dice score of 0.839. Finally, the third model used the Contrastive Unpaired Translation (CUT) method to generate fake hrT2, with 3D nnUNET for segmentation to attain a mean dice score of 0.829 \cite{third-winner}. Their approach mainly depends on doubling the number of images by generating augmented images with varying tumor intensities. Aside from the 2021 crossMoDA challenge, several approaches were followed for medical imaging domain adaptation, including weak supervision. In \cite{ws-reuben}, the authors applied a weak supervision methodology based on having partial annotations derived from scribbles on the target domain. They propose a technique that combines structured learning and co-segmentation to segment VS on T2 scans (target domain) from T1 scans (source domain). They achieved a dice score of 0.83 on the target domain

\section{Methods}
Our work consists of two public frameworks: Contrastive Unpaired Translation (CUT) \cite{park2020cut} for transferring ceT1 to hrT2, and nnUNet \cite{nnUNet} for segmentation. All the following work is implemented using PyTorch 1.11 and with the same mathematical formulation proposed in the CUT and nnUNet papers.
\subsection{Data}
The dataset contains 210 ceT1 MRI scans, 210 hrT2 MRI scans for training, and 64 hrT2 MRI scans for validation \cite{dataset}. This dataset is an addition to the publicly available Vestibular-Schwannoma-SEG dataset, a part of The Cancer Imaging Archive (TCIA), that was manually segmented \cite{crossMoDA}. The ceT1 and hrT2 scans are unpaired, and the segmentation masks are only provided for the ceT1 scans. The tumor is on one side of the brain, thus, only one of the cochleas would experience the pressure caused by it. Regardless, the segmentation masks include both the cochleas, as it was found that segmenting the two increases the performance of the models \cite{crossMoDA}. Image acquisition happened at two institutes in two locations: London and Tilburg. The testing set is made of hrT2 and is not publicly available. Evaluation of the testing set is made by the challenge organizers using the participant's submitted Docker container.
\subsection{Pre-processing}
The provided 3D scans differ in size and voxel spacing. Thus, we resampled the images into an isotropic resolution of 1 mm\textsuperscript{3}. During resampling, the interpolation techniques used are a third order b-spline for the images and nearest neighbor for the labels. Based on each image dimension, they were either cropped or padded in the $xy$-plane to 256 $\times$ 256, with a varying number of slices per image. The scans were normalized on a 3D basis, which led to better results during the domain translation phase. Previous techniques included normalization per slice. Since the slices have different ranges of pixel intensities, 2D normalization resulted in inconsistencies that decreased the quality of the generated hrT2 images during the domain translation phase. We note that domain adaptation processes that rely on generative algorithms are sensitive to the pre-processing techniques, as these results are essentially the input to the segmentation task.  
\subsection{Domain Translation}
\begin{figure}[h]
    \centering
    \includegraphics[width=0.8\textwidth]{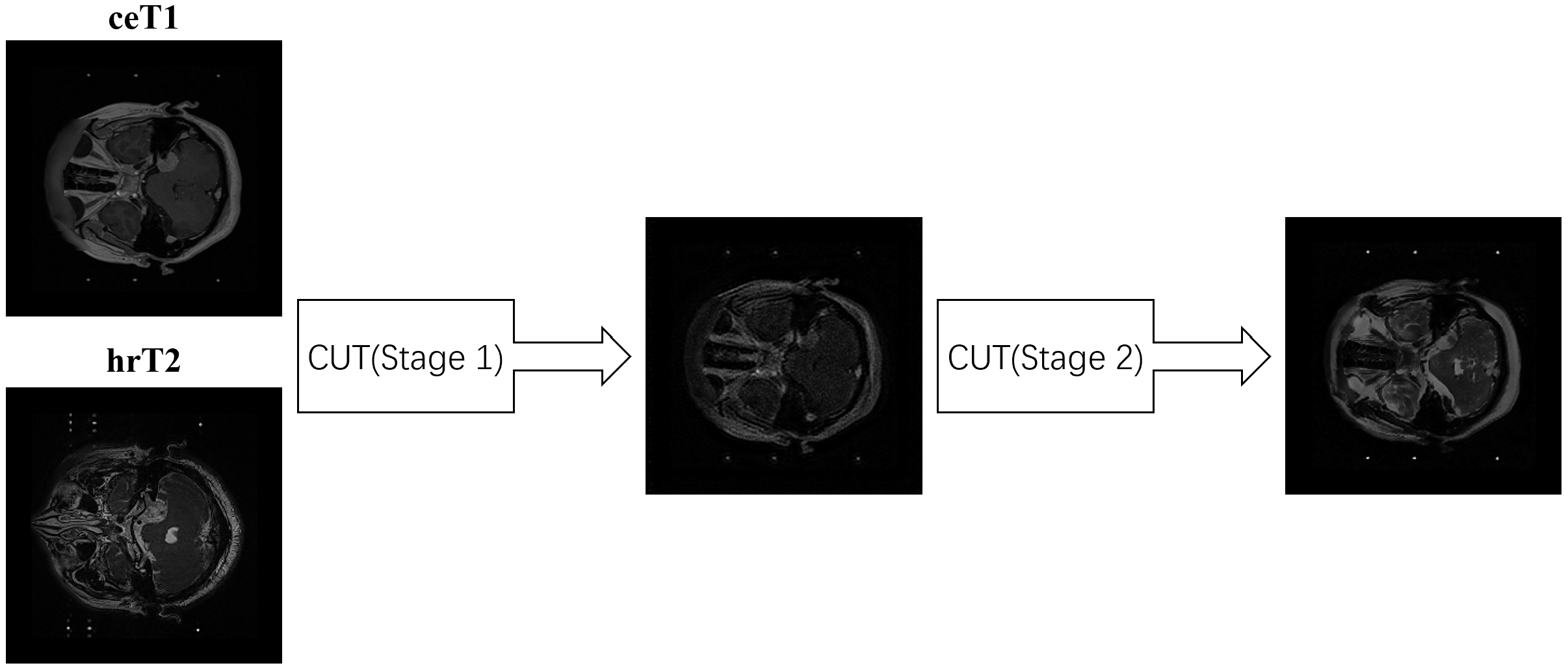}
    \caption{Overview of our CUT process. Dividing the training of CUT into two parts can speed up the training and avoid the discriminator being too strong. Keeping the batch size as 1 yields the same result.}
    \label{fig: cut_process}
\end{figure}

Our domain translation work is mainly based on CUT \cite{park2020cut} (based on CycleGAN without bijection request) framework to transfer ceT1 images to hrT2 images. In the original CUT paper, ResNet was used as the generator network and a multi-layer perceptron as the discriminator network. In our work, we used StyleGAN2 backbone \cite{stylegan2} instead, as it showed more promising results in the literature. Also, the authors trained the CUT for $N$ number of epochs with a fixed learning rate, followed by another $N$ epochs where the learning rate linearly decays to zero. Similarly, we applied this approach during our domain translation phase.\\ We trained the CUT model in two stages to speed up the training and avoid the discriminator over-shadowing the generator which could make it generate poorly representative hrT2 images. In the first stage, we set a batch size of 32 and a learning rate of 0.001 for the first 50 epochs. Then, for the second 50 epochs, the learning rate was linearly decaying to reach $0$ at the final epoch. We used the Fréchet Inception Distance (FID) \cite{Seitzer2020FID} as a metric to evaluate the proximity of the generated image to the target domain.
In the second stage, we re-initialized the discriminator with random weights. This makes it more challenging for the discriminator to distinguish the views, allowing the generator to learn a better representation of hrT2 images. If the discriminator was not re-initialized, its effect will dominate, and the generator will produce low-quality hrT2 scans. We trained the CUT for 5 epochs with a learning rate of 0.001, followed by another 5 epochs where the learning rate experienced a linear decay to finally reach 0. The method is summarized in Table \ref{tab:cut}. In addition, Figure \ref{fig: cut_process} shows 2D slices of a scan after passing through stage 1 and clarifies how it gets clearer after stage 2. We used these pseudo hrT2 images as the training dataset of the segmentation network.

\begin{table*}[ht!]
\caption{Presented is the methodology followed during the domain translation phase. The table shows the settings of the two training stages of the CUT. BS refers to batch size, while LR refers to the learning rate.}
\begin{center}
\scalebox{0.68}{
\begin{tabular}{|c|c|c|c|c|c|c|}
\hline
Stage&BS&LR&Generator network baseline&Discriminator network baseline&Epochs&Epochs to decay\\
\hline
1& 32 & 0.001&StyleGAN2 instead of ResNet9&StyleGAN instead of MLP&50&50\\
\hline
2& 1 & 0.0002&StyleGAN2 instead of ResNet9&StyleGAN instead of MLP&5&5\\
\hline
    \end{tabular}
}
\end{center}
    \label{tab:cut}
\end{table*}

\begin{figure}
    \centering
    \includegraphics[width=0.7\linewidth]{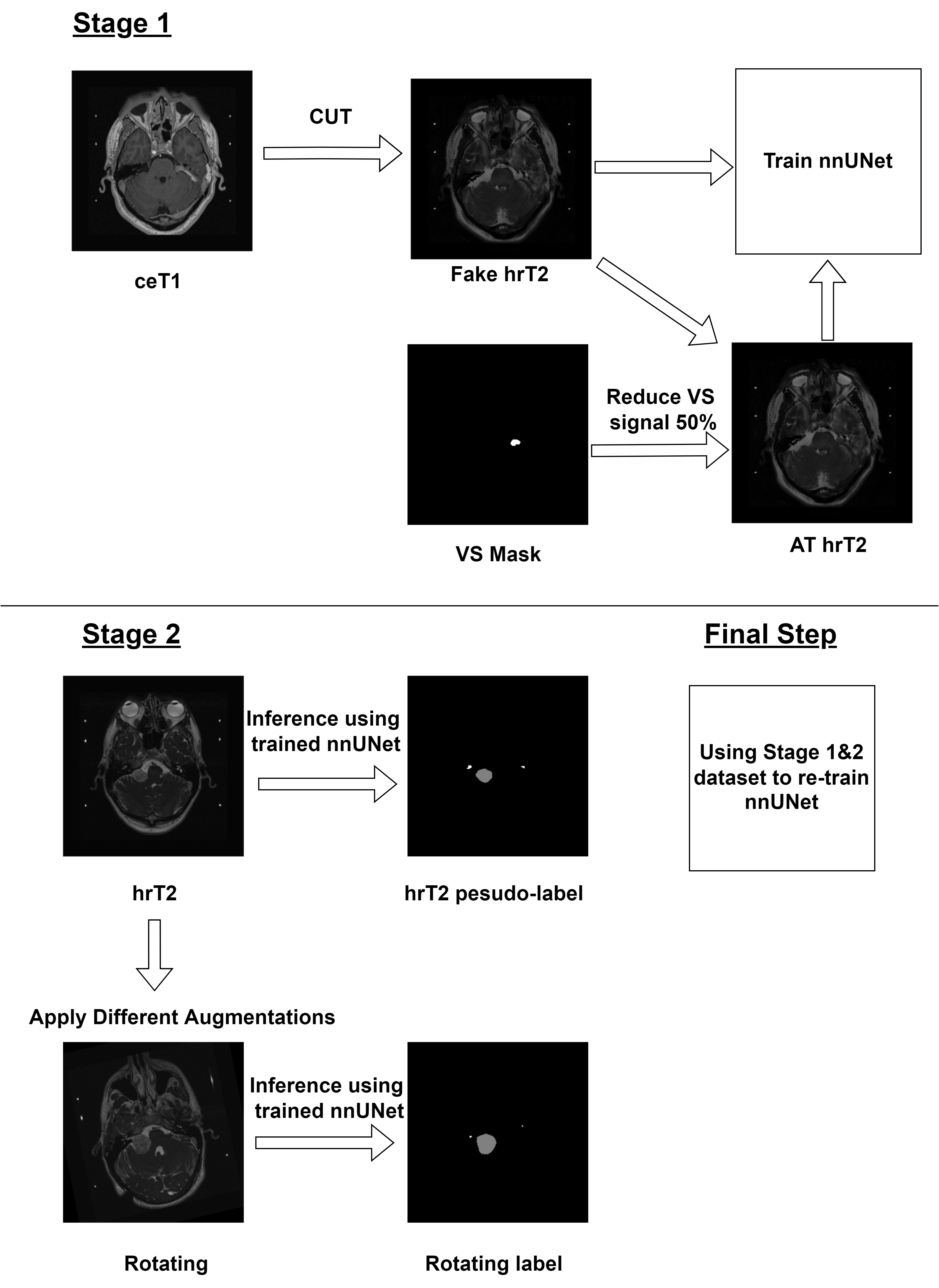}
    \caption{Overview of the model: Stage 1 includes generating the fake hrT2 images, using them to create the tumor augmented version, and training using fake hrT2 images and the tumor augmented version. Stage 2 involves applying different augmentations to the real hrT2 images and producing their pseudo-labels from the model at stage 1.}
    \label{fig: model_arch}
\end{figure}
\subsection{Segmentation} \label{method:segmenation}
For the segmentation task, we used the nnUNet framework with the 3D full-resolution U-Net configuration. Following the approach of \cite{third-winner}, we generated the augmented fake hrT2 images by reducing the tumor signal by 50\%, naming them the AT dataset. By that, we had a total of 420 training images. We applied five-fold cross validation using the default nnUNetTrainerV2 which combines two losses: cross entropy and dice, which are
\begin{equation}
    \mathcal{L}_{dice} = \frac{-2}{C}\sum_{c\in C} \frac{\sum_{i}p_i^c g_i^c +
    \epsilon}{\sum_{i}p_i^c + \sum_{i}g_i^c + \epsilon}
    \label{dice}
\end{equation}
\begin{equation}
    \mathcal{L}_{CE} =  \sum_{i}-g_i\log(p_i) - (1-g_i)log(1-p_i)
\end{equation}
where $i$ refers to the voxel, $p$ is the predicted mask, $g$ is the ground truth, $C$ is the number of classes, and $\epsilon$ is the smoothing parameter. In the default setting, $\epsilon$ is set to 1. We also used the nnUNet variant that applies the non-smooth dice loss instead of the regular one, having $\epsilon =0$ in Equation \ref{dice}. The non-smooth dice loss proved its effectiveness when the structure of interest is small relative to the scan size, such as the cochlea. We still noticed a gap between the performance on the training and validation datasets. Therefore, to alleviate the generalizability of the model, we used a nnUNet variant which applies multiple augmentation techniques to the training data. Since \cite{first-winner} showed that generating pseudo-labels of real hrT2 images to train on increases the dice score, we followed a similar approach. Thus, our models were then trained on a total of 630 images. \\
For all the nnUNet variants used in this work, instance normalization was used. Following \cite{se}, we applied squeeze-excitation normalization that enhances channel interdependencies. The SE blocks include a global average pooling layer, two fully connected layers (FC), and activation functions. The aim of the pooling layer is to squeeze the channel information into one value, while the FC layers learn the non-linear dependencies between the channels. The model involved a reduction rate ratio of 2 and a ReLU activation function. Other than the normalization approach, all settings were kept the same as in the default nnUNet. \\
For the final model, we made it depend on augmentations as a weak supervision approach as shown in Figure \ref{fig: model_arch}. We applied eight types of augmentations on the real hrT2 dataset and predicted their pseudo-labels. The augmentation techniques are random rotation of up to 20 degrees, adding noise, scaling and translating, changing the contrast, and flipping on the three axes. As a result, we had 2100 images with a majority of real hrT2 images. This approach allows our model to better learn the features specific to the hrT2 modality and makes it less prone to the discrepancy arising from the GAN used. All the nnUNet variants apply the deep supervision approach for the loss functions which helps with the gradient vanishing problems for deep networks. Lastly, we ensembled two models involving: the augmentations variant with the combined loss, and the augmentations variant with the non-smooth dice loss. When predicting on an unseen case, we post-process the mask by finding the largest connected component of the VS. This is because the model tends to segment it on both sides of the brain while it is only located on one.
\section{Results}
As for the CUT model, the first stage achieved an FID of 70.37, while the second stage improved the model, reaching an FID of 51.3. Regarding the nnUNet implementation, Table \ref{tab:results} presents some of the results acquired from the different settings on the validation dataset. The baseline model included the fake hrT2 images with their tumor augmented version. The mean dice score achieved is 0.73, with low performance on segmenting the cochleas. Then, two models were trained after producing the pseudo-labels of the original hrT2 images, giving almost similar results: a mean dice score of 0.76. We then experimented by replacing the instance normalization in the default nnUNet with a squeeze-excitation normalization. However, obtaining a mean dice score of 0.73, we noticed no significant improvement using SE normalization. Thus, we used instance normalization in the rest of the experiments.\\
After that, we experimented with various augmentation techniques on the training data. Table \ref{tab:results} describes the performance of the augmentations variant models. Relatively, tested on the validation dataset, our analysis concludes that the two models with augmentations during training gave the best results. As a consequence, we produced augmented real T2 images with their pseudo-labels and ran the final ensemble model described in Section \ref{method:segmenation}. The ensemble model achieved a mean dice score of 0.77 on the validation set, with a considerable improvement on the ASSD of the cochleas, reaching 0.37. As we can see from Table \ref{tab:results}, this model has a lower variance compared to the other proposed models, which could be due to its learning to focus on the regions that are shared with the augmented pseudo-label. \\
Since the testing dataset is not made public, our performance metrics were obtained by the challenge organizers. Our model achieved a mean dice score of 0.78 and an ASSD of 0.46. As to the VS, the dice score is 0.83 and the ASSD is 0.56. The cochleas have a dice score of 0.74 and an ASSD of 0.35. 
\begin{table*}[ht!]
\caption{Results from different nnUNet variants. The best model is indicated in bold and represents an ensemble of two experiments: nnUNet with augmentations and non-smooth dice loss, and nnUnet with augmentations and combined loss.``Fake hrT2" are generated from the CUT, ``AT" represents images where tumor signal is reduced by 50\%, ``aug var" describes the augmentations variant, and ``pseudo-labels" are inferences of the real hrT2 scans.}
\begin{center}
\scalebox{0.68}{
\begin{tabular}{|c|c|c|c|c|c|c|c|}
\hline
Model&Dataset&Epoch&Score $\uparrow$&VS Dice $\uparrow$&VS ASSD $\downarrow$&Cochlea Dice $\uparrow$&Cochlea ASSD $\downarrow$\\
\hline
nnUNetTrainerV2& fake hrT2 + AT& 300 & 0.73$\pm$0.08&0.79$\pm$0.13&0.73$\pm$0.43&0.68$\pm$0.07&0.63$\pm$1.80\\
\hline
nnUNetTrainerV2 &\makecell{fake hrT2 + AT \\+ pseudo-labels} & 800 &0.76$\pm$0.07&0.81$\pm$0.11&1.14$\pm$1.71&0.71$\pm$0.06&0.60$\pm$1.80\\
\hline
\makecell{nnUNetTrainerV2 \\ + non-smooth dice}& \makecell{fake hrT2  + AT\\+ pseudo-labels} & 800 &0.76$\pm$0.07&0.81$\pm$0.11&1.28$\pm$1.84&0.71$\pm$0.06&0.59$\pm$1.80\\
\hline
\makecell{nnUNetTrainerV2 \\ + SE normalization}& fake hrT2  + AT & 750 &0.73$\pm$0.09 &	0.76$\pm$0.15&1.74$\pm$2.50&0.69$\pm$ 0.07&0.62$\pm$1.79\\
\hline
\makecell{nnUNetTrainerV2 \\+ aug var} &fake hrT2  + AT& 500 & 0.75$\pm$0.07&0.80$\pm$0.11&0.69$\pm$0.38&0.70$\pm$0.06&0.62$\pm$1.80\\
\hline
\makecell{nnUNetTrainerV2 \\aug var + non-smooth dice} & fake hrT2  + AT& 500 & 0.75$\pm$0.08&0.79$\pm$0.13&0.71$\pm$0.41&0.71$\pm$0.06&0.61$\pm$1.80\\
\hline
\makecell{Ensemble of \\ (nnUNetTrainerV2 aug var \\ + non-smooth dice) \\ \& (nnUNetTrainerV2 \\aug var)} & \makecell{fake hrT2  + AT \\+ pseudo-labels} & 1000 &  \textbf{0.77$\pm$0.06}& \textbf{0.82$\pm$0.09}&\textbf{0.61$\pm$0.27}&\textbf{0.72$\pm$0.06}&\textbf{0.37$\pm$0.18}\\
\hline
    \end{tabular}}
\end{center}
    \label{tab:results}
\end{table*}
\section{Qualitative Analysis}
\begin{figure}[h]
     \centering
     \begin{subfigure}[b]{0.4\textwidth}
         \centering
         \includegraphics[width=\textwidth]{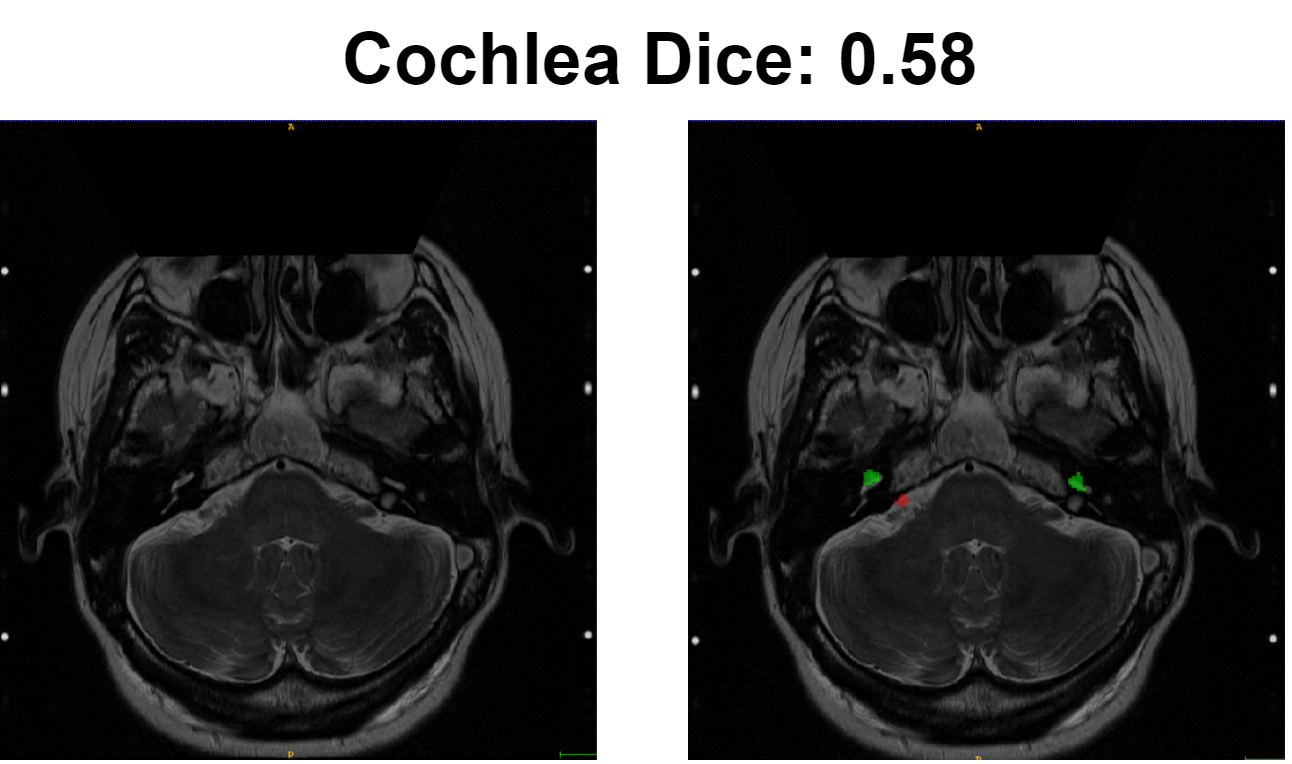}
         \caption{Worst Cochlea Score}
         \label{Worst Cochlea Score}
     \end{subfigure}
     \hfill
     \begin{subfigure}[b]{0.4\textwidth}
         \centering
         \includegraphics[width=\textwidth]{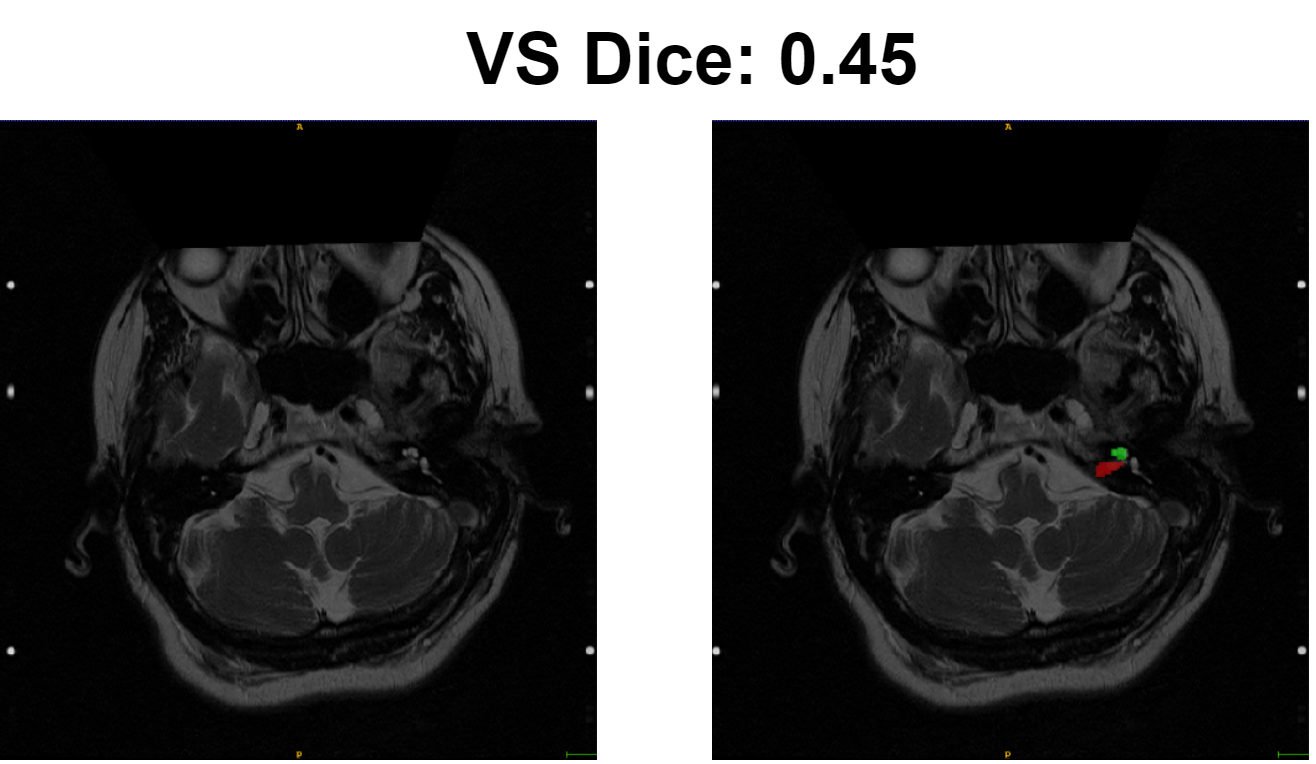}
         \caption{Worst VS Score}
         \label{Worst VS Score}
     \end{subfigure}
     \hfill
    \par\bigskip
    \par\bigskip
      \centering
     \begin{subfigure}[b]{0.4\textwidth}
         \centering
         \includegraphics[width=\textwidth]{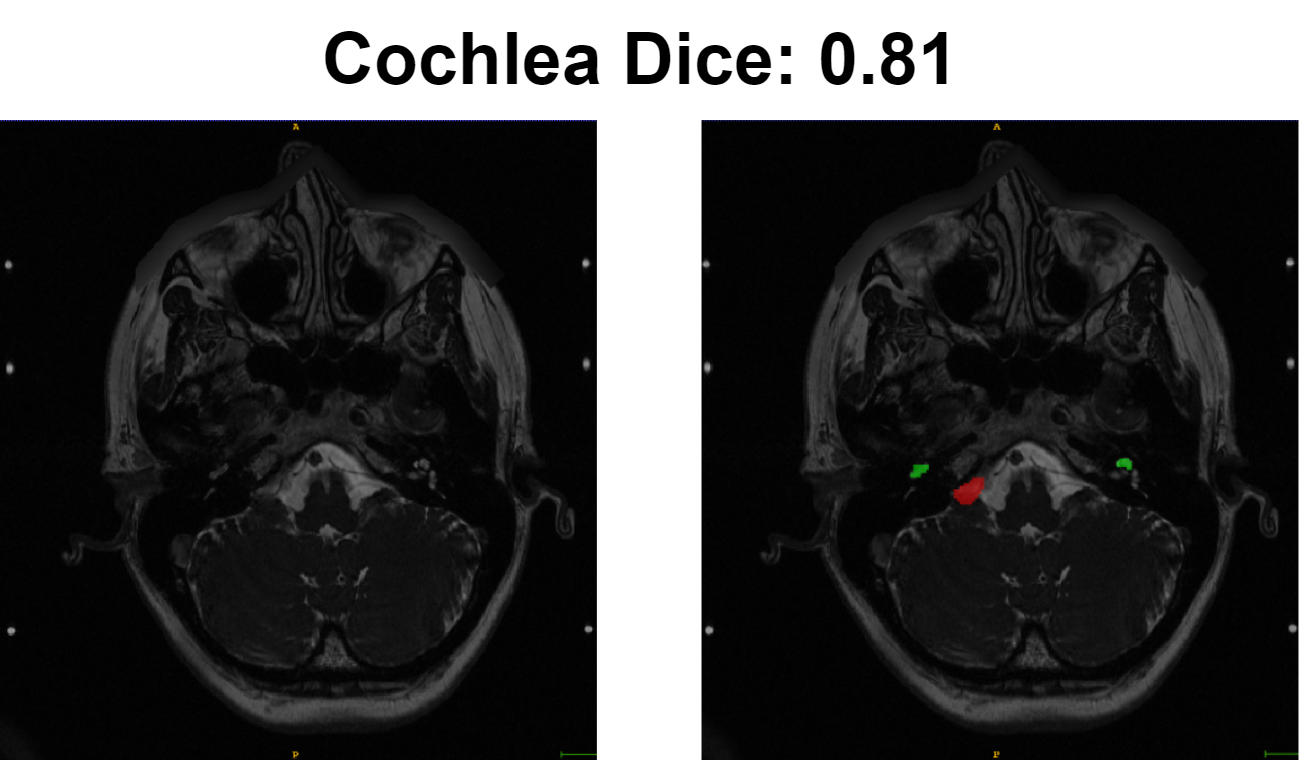}
         \caption{Best Cochlea Score}
         \label{Best Cochlea Score}
     \end{subfigure}
     \hfill
     \begin{subfigure}[b]{0.4\textwidth}
         \centering
         \includegraphics[width=\textwidth]{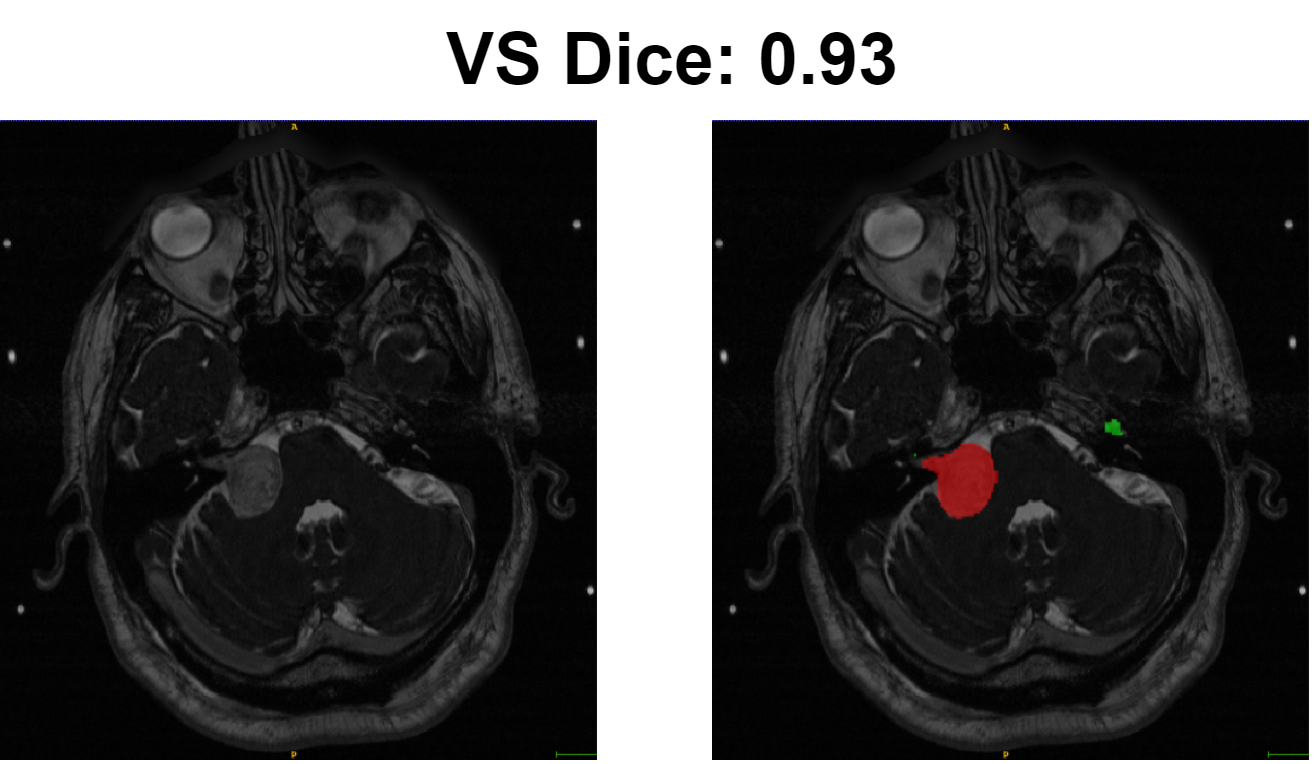}
         \caption{Best VS Score}
         \label{Best VS Score}
     \end{subfigure}
    \caption{The best and the worst segmentation according to the dice score for Cochlea (in green) and VS (in red).}
    \label{fig: quality_images}
\end{figure}

Our best and worst segmentation results of our final model on the validation set are shown in Figure \ref{fig: quality_images}. The final model is an ensemble of two networks: nnUNet with augmentations and non-smooth dice loss, and nnUNet with augmentations and combined loss. It is still possible to discuss a few points even though we do not have the ground truth mask at hand. As observed in Figure \ref{Worst Cochlea Score}, the model missed parts of the cochleas during segmentation and over-segmented the background. On the other hand, in the best case scenario for the cochlea in Figure \ref{Best Cochlea Score}, the model over-segments the cochlea region. Based on the two VS cases in Figures \ref{Worst VS Score} and \ref{Best VS Score}, we can see that the model can segment clear tumors well, while dark tumors are more difficult to segment.

\section{Discussion}
We developed a deep learning algorithm that follows a weak supervision approach to segment two brain structures in the hrT2 modality. The model uses a GAN to generate fake hrT2 images from the ceT1 scans, then these hrT2 images are trained using an nnUNet. After that, the fake hrT2 images are augmented and the nnUNet is used to predict the pseudo-labels of the augmented images. Finally, the model is retrained using the real hrT2 images and the latter augmented hrT2 images as the training dataset, and is validated on the validation dataset provided from the challenge organizers. The applied experiments vary in the loss function, datasets, and augmentations. We observed that the implementation of an accurate GAN model plays a huge role in the success of the unsupervised DA model. Given that the model is predicting an unseen modality, the more generalizable it is, the better. Thus, even if the augmentations increase the size of the training data, it offers the model an optimal chance to diversify the features learned from the target domain. We noticed that the model is more consistent in predicting the cochlea, but less accurate, relative to segmenting the VS. This may be due to the cochlea's size being small in relation to the entire image, but has a fixed location across the patients' scans. However, the VS prediction is not consistent because the tumor intensity, location, and size vary by patient. Moreover, the ASSD metric improved significantly in the final model compared to the improvement noticed in the dice score, especially for the cochleas. We associate it with the different augmentations applied during training, which led to a better study of the overall shape of the structures in different orientations. The ASSD considers the distance between the boundaries of the ground truth and the predicted mask which is more sensitive to outliers in prediction. Thus, we can conclude that the improvement in ASSD indicates that the final model is more robust than the previous models.
\section{Conclusion}
The application of domain adaptation enables ML models to aid in cases similar to the VS disease, where there is a growing interest in using hrT2 images, but not enough data. Both the settings of the GAN used and the segmentation network impacts the efficiency, especially for small-size structures, such as the cochleas. In some cases, weak supervision approaches may be computationally expensive. Regardless, they enable the model to learn different properties and views of the target domain scans. Thus, they maximize the learning of features and increase generalizability. Since the augmented hrT2 scans are predicted from the initial nnUNet network, improving it would significantly improve the quality of the prediction. Therefore, further work includes higher performing segmentation models, especially on the cochleas, in order to be combined with our weak supervision approach.

\bibliography{ref} 
\bibliographystyle{splncs04}

\end{document}